\documentclass{article}
\usepackage{fullpage}

\usepackage[affil-it]{authblk}

\usepackage[super,sort&compress]{natbib}
\usepackage{amsmath}
\usepackage{amsfonts}
\usepackage{graphicx}

\usepackage{hyperref}
\usepackage{wrapfig}
\usepackage{subcaption}
\makeatletter
\renewcommand\@maketitle{%
\hfill
\begin{minipage}{0.95\textwidth}
\vskip 2em
\let\footnote\thanks 
{\LARGE \@title \par }
\vskip 1.5em
{\large \@author \par}
\end{minipage}
\vskip 1em \par
}
\makeatother

\begin{document}

\title{ Range separation: the divide between local structures and field theories}

%\author{The Author}
%\affiliation{The Author's Institution}
%\email{email@email.edu}
\author{ David M. Rogers}
\affil{ University of South Florida, 4202 E. Fowler Ave., CHE 205, Tampa, FL 33620}
%\date{\today}
%\keywords{electronic structure, liquid state structure, density functional theory, Bayes' theorem, vapor interface, molecular dynamics}

%\singlespacing

\maketitle
\thanks{ Electronic address: \texttt{davidrogers@usf.edu}}

\begin{abstract}
This work presents parallel histories of the development of two modern theories of
condensed matter: the theory of electron structure in quantum
mechanics, and the theory of liquid structure in statistical mechanics.
Comparison shows that key revelations in both are not only
remarkably similar, but even follow along a common thread of controversy that
marks progress from antiquity through to the present.
This theme appears as a creative tension between two 
competing philosophies, that of short range structure 
(atomistic models) on the one hand, and long range structure 
(continuum or density functional models) on the other.
The timeline and technical content are designed to
build up a set of key relations as guideposts 
for using density functional
theories together with atomistic simulation. \\

\end{abstract}

% field: physical chemistry, mathematical physics, computational chemistry
\vspace{3em}

  Many of the most important scientific theories were forged out of controversy -- like particles vs. waves, for which Democritus claimed (with his teacher, Leucippus of 5th century BC) that all things, including the soul, were made of particles, while Aristotle held to the Greek notion that there were continuous distributions of four or five elements.\cite{atom}  It is telling to note that Aristotle's objection was strongly biased by his notion that the continuum theory was elegant and beautiful, and does not require any regions of vacuum.  In addition, his conception of kinetic equations were first order -- like Brownian motion, Navier-Stokes, or the Dirac equation, but not second order like Newton's or Schr\"{o}dinger's.  Newton sided with Democritus.  In 1738, Daniel Bernoulli first explained thermodynamic pressure using a model of independent atomic collisions.  That theory was not scheduled to be widely adopted until the caloric theory (which postulated conservation of heat) was overthrown by James Joule in the 1850s.  Wilhelm Ostwald was famously stubborn for refusing to accept the atomic nature of matter until the early 1900s, after Einstein's theory of Brownian motion was confirmed by Jean Perrin's experiment.

  The working out of gas dynamics by Maxwell and Boltzmann in the 1860s depended critically on switching between a physical picture of a 2-atom collision and a continuum picture of a probability distribution over particle velocities and locations (Fig.~\ref{f:hybrid}a).  Collision events drawn at random from a Boltzmann distribution were useful for predicting pressures and reaction rates.  Whether that distribution represented a probability or an actual average over a well-enough defined physical system was left open to interpretation.  Five decades later, Gibbs would argue with Ehrenfest\cite{pehre59} over this issue.
Gibbs seemed to understand the continuous phase space density as {\em any} probability distribution
that met the requirements of stationarity under time evolution.  An observer with no means of
gathering further information would have to accept it as representing reality.
Ehrenfest argued that a well-defined physical system is exact, mechanical, and objective.
The controversy was only resolved by the advent of the age of computation,\cite{bninh17} since we forgot about it.
Three decades on, the physicist Jaynes championed the (subjective)
maximum entropy viewpoint,\cite{ejayn79} while mathematicians like Sinai and Ruelle\cite{druel03,ggall08,svara08,htouc09}
moved to do away with the whole subjectivity business
by using only exact dynamical systems as starting assumptions.
%(or at least restrict it to uniform distributions\cite{ejayn80b}).

\begin{table}
{\centering 
\small
\begin{tabular}{p{1in}llp{1in}}
& SR/Discrete & LR/Continuous & \\
\hline\hline
(Democritus) & atoms & elements & (Aristotle) \\
(Ehrenfest) & microstate & ensemble & (Gibbs) \\
(Einstein) & particle & wave & (Ostwald) \\
(Boltzmann) & distribution function & 1-body probability density & (Jaynes) \\
\hline
(Wein) & $n(\nu)$ & $\nu^2 d\nu$ & (Rayleigh-Jeans) \\
& $\hat n(r,p)$ & $n(r), V^\text{ext}(r)$ & \\
&  & Jellium & (Sommerfeld) \\
(Mott) & insulator & conductor & (Pauli) \\
(Hartree-Fock) & Slater determinant & Electron density & (Hohenberg-Kohn-Sham) \\
(Born-Oppenheimer) & nucleii & electrons & \\
& correlation hole & polarization response & \\
\hline
(Bohm-Pines) & \multicolumn{2}{c}{$\xleftarrow{\phantom{very long text}}$ Quasiparticle}  & \\
& \multicolumn{2}{c}{Phonon $\xrightarrow{\phantom{very long text}}$}  & \\
& \multicolumn{2}{c}{$\xleftarrow{\phantom{very long text}}$ Cooper Pair}  & \\
& \multicolumn{2}{c}{Hybrid DFT $\xrightarrow{\phantom{very long text}}$}  & \\
%& \multicolumn{2}{c}{$\xleftarrow{\phantom{very long text}}$ Exact Exchange} & \\
\hline
\end{tabular}
\caption{Contrasting long-range (LR) and short-range (SR) ideas showing stages of debate over atoms and
electrons (top two sections), along with concepts from hybrid theories (lower section).}\label{t:range}
}
%\vspace{-0.5em}
\end{table}

  Maxwell described light propagation by filling the continuum with `idler wheels,' and the resulting partial differential equations inspired much of 20th century mathematics. % -- particularly the Sturm-Liouville eigenvalue and Green's function methods.
Planck saw his own condition on quantized transfer of light energy as a
regrettable, but necessary refinement of Maxwell's theory.
Planck believed so strongly in that theory that he at first
rejected Einstein's 1905 concept of the photon.\cite{stoul67}
It was also five decades later, around 1955, when a field theory of the electron
(quantum electrodynamics) was gaining acceptance from
precise calculations of experimental details like the gyromagnetic ratio, radiation-field drag (spontaneous emission)
and the Lamb shift.  This quantum field theory is not a completely smooth
continuum, since it incorporates particles using `second quantization.'
It understands particles as wavelike disturbances that
pop in and out of existence in an otherwise continuous field.
The technical foundations of that theory are derived by `path-integrals'
over all possible motions of Maxwell's idler wheels.
As a consequence, infinities characterize the theory,\cite{ejayn90}
so that the mathematical status of many path integrals
is still not settled\cite{fdyso72} except in the Gaussian case,\cite{dwitt72,mmizr78} and
where time-sliced limits are well-behaved.\cite{hklein09}

  This article discusses some well-known historical developments in
the theory of electronic and liquid structure.
As its topic is physical chemistry, this history vacillates without warning between
experimental facts and technical details of the mathematical models conjured to describe them.
The topics, outlined in Table~\ref{t:range}, have been chosen specifically to
highlight the debate between local structural and field theoretical models.
Note that we have also presented the two topics in an idiosyncratic
way to highlight their similarities.  Differences between electronic and liquid
structure theories are easy to find.
By the nature of this type of article, we could not hope to be comprehensive.
There has not been space to include many significant historical works,
while it is likely several offshoots and recent developments have been unknowingly overlooked.
Both histories trace their roots to the Herapath/Maxwell/Boltzmann conception
of a continuous density (or probability distribution) of discrete molecules,
and both remain active research areas that are
even in communication on several points.
We will find that, like Democritus and Aristotle, not only are there are strong opinions
on both sides, but progress continues to be made by researchers regardless of
whether they adopt discrete or continuum worldviews.

\section*{Electronic Structure Theories}

\begin{wrapfigure}{l}{0.5\textwidth}
{\centering
\includegraphics[width=0.5\textwidth]{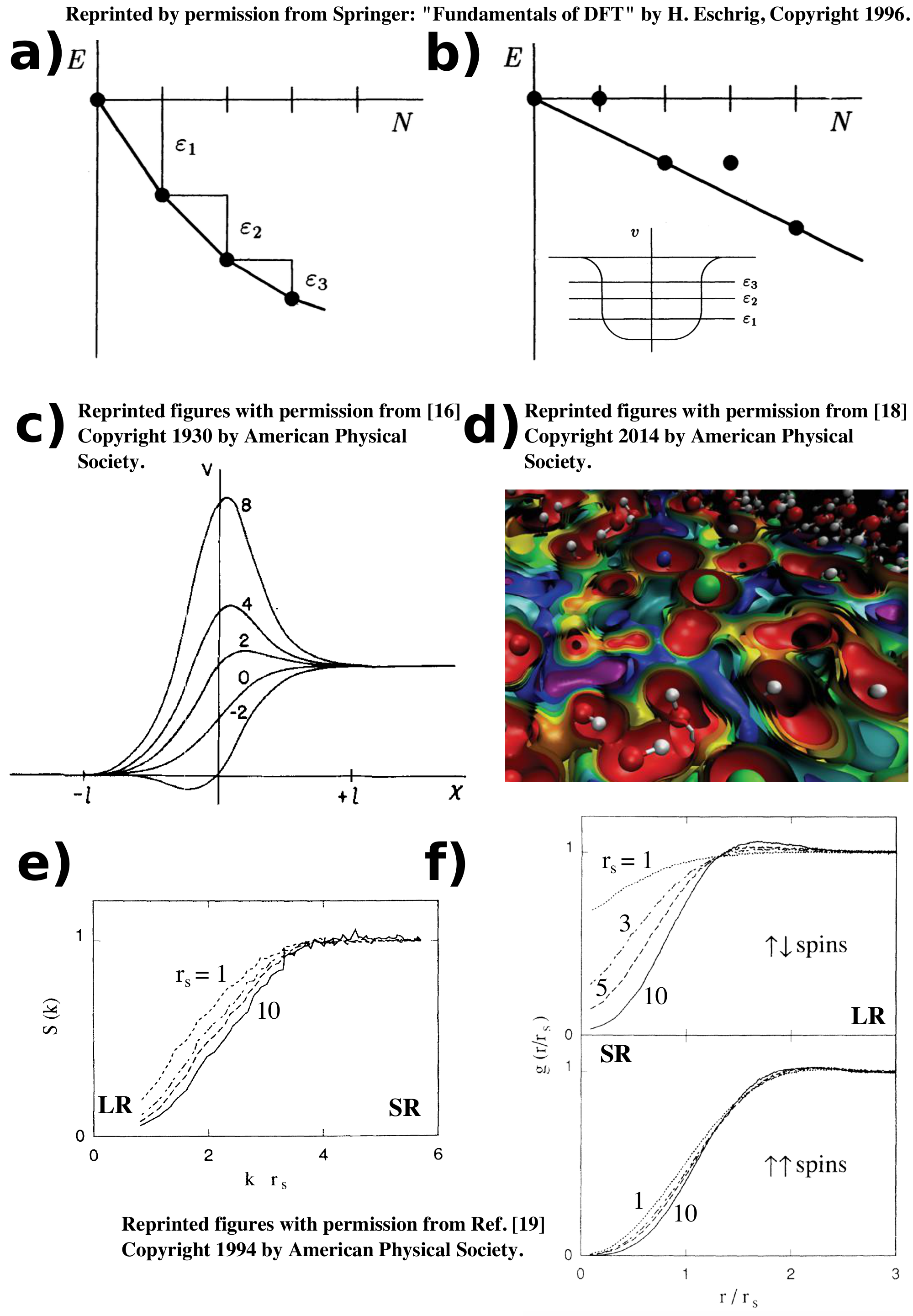}
%\caption{
%(a) Adapted by permission from Springer: Springer, Fundamentals of Density Functional Theory by H. Eschrig, 1996.}
\caption{Long-range (left) and short-range (right) theories of electronic structure.
(a) and (b) show free energy vs. electron number for a potential well.\cite{eschrigbook}
(c) shows `Epstein' profile of
dielectric response\cite{cecka30,srice70} at a metal/vacuum interface.
Numbers for each curve give the surface/bulk conductivity ratio.
(d) shows surfaces of constant voltage at a water/vacuum interface,\cite{bsell14}
(e) and (f) show the correlation function of jellium from accurate calculations.\cite{gorti94}}\label{f:ueg}
}
\vspace{-1.5em}
\end{wrapfigure}

% Jellium and its response function
  Between the lines of the history above, we find Bose's famous 1924 Z. Physik paper describing the statistics of bosons,
which Einstein noted `also yields the quantum theory of the ideal gas,' and the Thomas-Fermi theory of 1927-28 for
a gas of electrons under a fixed applied voltage.  Their basic conception was to model the 6-dimensional space of particle
locations, $r$ and momenta, $p$ with the volume element,
\begin{align}
g(p') dp' &= dp' \int \delta(|p|-p') h^{-3}\, dr^3 \, dp^3 \notag \\
  &= 4\pi V h^{-3} p'^2 dp'
\end{align}
Using $p' = h\nu/c$ for photons of frequency $\nu$ provides $g(\nu)$, the number of
available states for photons near frequency $\nu$.  Applying Bose counting statistics to $n(\nu)$
photons occupying $2 g(\nu)$ possible states for each frequency gives Bose's derivation of Planck's law.
In the Thomas-Fermi (TF) model, $p'$ is electron momentum.  Applying Fermi statistics
to the occupancy number $N = 2 \int^{\hbar k_F}_0 g(k\hbar) d(k\hbar)$
now gives a Fermi distribution for an ideal gas of electrons under a constant external potential (electrostatic voltage).
In both cases the number of states is doubled -- counting 2 polarizations for photons or 2 spin states for electrons.
The result of the first procedure is a free energy expression for the vacuum.
The result of the second is a free energy for electrons under a constant voltage. %, $\phi$.
%putting one electron in every state up to maximum momentum $p'^2 < 2 m_e e(\phi - E_0)$
%($m_e$ is the electron mass and $E_0$ sets the average electron energy).

% Step 1: completely explain n(r) and V(r) in the Thomas-Fermi theory
  This idea of a gas with uniform properties uses a long-range field to guess at local structure.
Quantitatively, if the voltage at point $r$ is $\phi(r)$, then the theory predicts
electrons will fill states up to maximum momentum of $k_F = \sqrt{2 m_e e_0 \phi(r)}/\hbar$,
(where the kinetic energy is $E_F = \hbar^2 k_F^2/2m_e$ and $e_0$ is the electron charge)
so the local density is,
\begin{equation}
n(r) = k_F^3/3\pi^2. \label{e:TF}
\end{equation}
The resulting model is then usually found to predict long-range properties of metals relatively well.
Fig.~\ref{f:ueg}a and~b show plots of free energy {\em vs} number of electrons
in an independent electron solution of the Schr\"{o}dinger equation for a well of positive potential.\cite{eschrigbook}
Panel~b shows a simple adaptation of that model where electrons
bind in pairs.  The states of the electrons in these exact solutions still represent momentum
levels, and are thus qualitatively very close to those of the Thomas-Fermi theory.

  The free electron gas evolved into the famous `jellium' model of electron motion
rather quickly, as can be seen by the earliest references in a discussion of that model
from the late 20th Century.\cite{nlang70}
The term jellium was coined by Conyers Herring in 1952 to describe the model
of a metal used by Ewald\cite{pewal52} and others consisting of a uniform background density of positive charge.
The electrons are therefore free to move about in gas-like motion.  At high density,
the electrons actually do act like a free gas, so it was possible to use the Thomas-Fermi
theory to qualitatively describe the electronic contribution to specific heat,
$C_v = \pi^2 k_B^2 T/2 E_F$, as well as the spin susceptibility
and width of the conduction band (after re-scaling the electron mass).\cite{pines3}
These are long-range properties arising from the collective motion of many electrons.
The predictions become poor for semi-metals and transition metals.
It also rather poorly described the cohesive energy of the metal itself.
Those cases fail because of the importance of short-range interactions
that a free electron theory just doesn't have.\cite{jslat35}
% Note: `the' response function of Jellium from Rice?
%Pines' book describes the process of adding a perturbation based on Hartree-Fock
% theory to add short-range corrections to the long-range behavior of the electrons above.
% The uniform electron gas (no neutralizing background, just subtract avg.)

  The contrast becomes important at interfaces, as is visible when comparing Fig.~\ref{f:ueg}c,d.
On the left is an early model of local charge density response due to placing an external
voltage at a point near a metal surface.  On the right is a map of
the local voltage for one surface configuration of an electrolyte solution
computed using an accurate quantum density functional theory.
Chloride ions are green, and sodium ions are blue.  Treating one of the
sodium ions as a test charge, the material response comes from
rearrangement of waters (red and white spheres) and Cl$^-$ ions
within a nuanced voltage field (colored surfaces).

\begin{figure}[t]
    \centering
    \begin{subfigure}[t]{0.48\textwidth}
        \includegraphics[width=\textwidth]{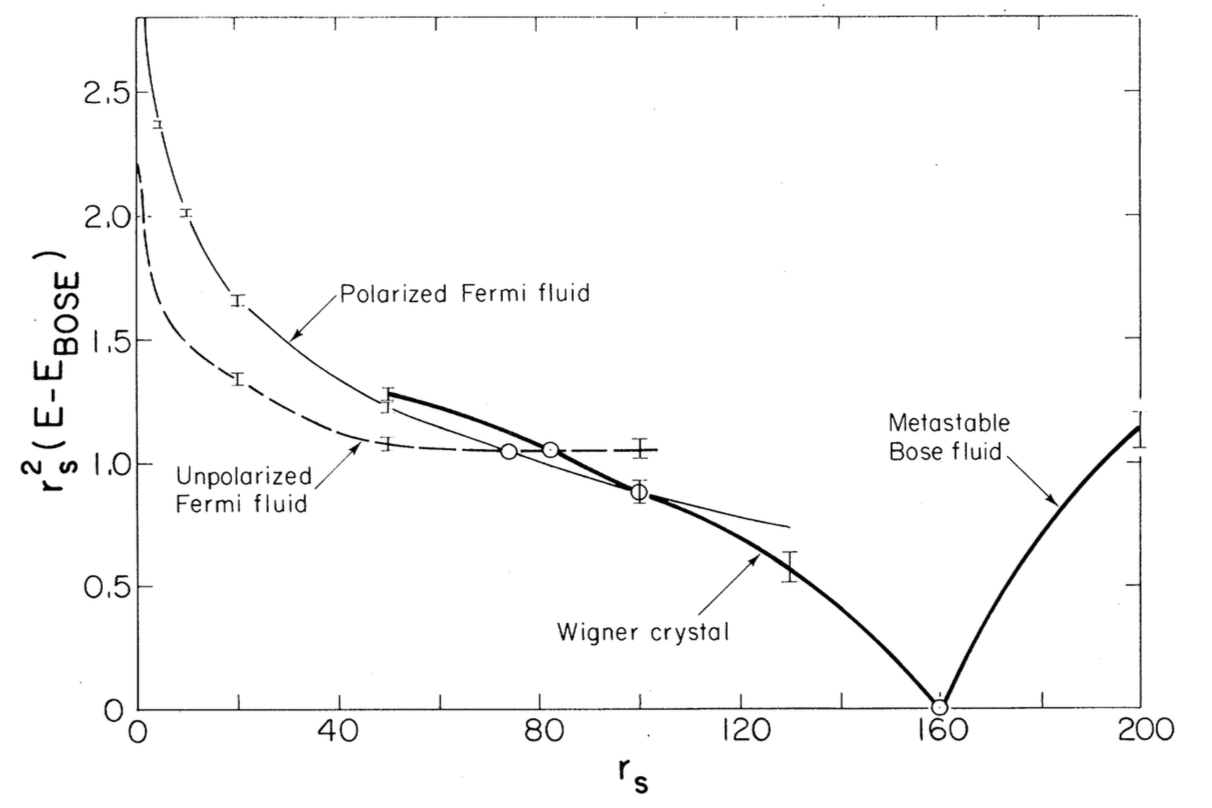}
        \caption{Ground state energy vs. density for the uniform electron gas.\cite{dcepe80}
        		     Four separate phases were observed (at zero temperature).
        		     Note that the density axis is reversed by the transformation $1/n = 4\pi r_s^3/3$.
		     Reprinted figure with permission from
		     \href{https://dx.doi.org/10.1103/PhysRevLett.45.566}{Ref.~\citenum{dcepe80}}.
		     Copyright 1980 by the American Physical Society.}
        \label{f:ceperly}
    \end{subfigure} \,
    \begin{subfigure}[t]{0.48\textwidth}
        \includegraphics[width=\textwidth]{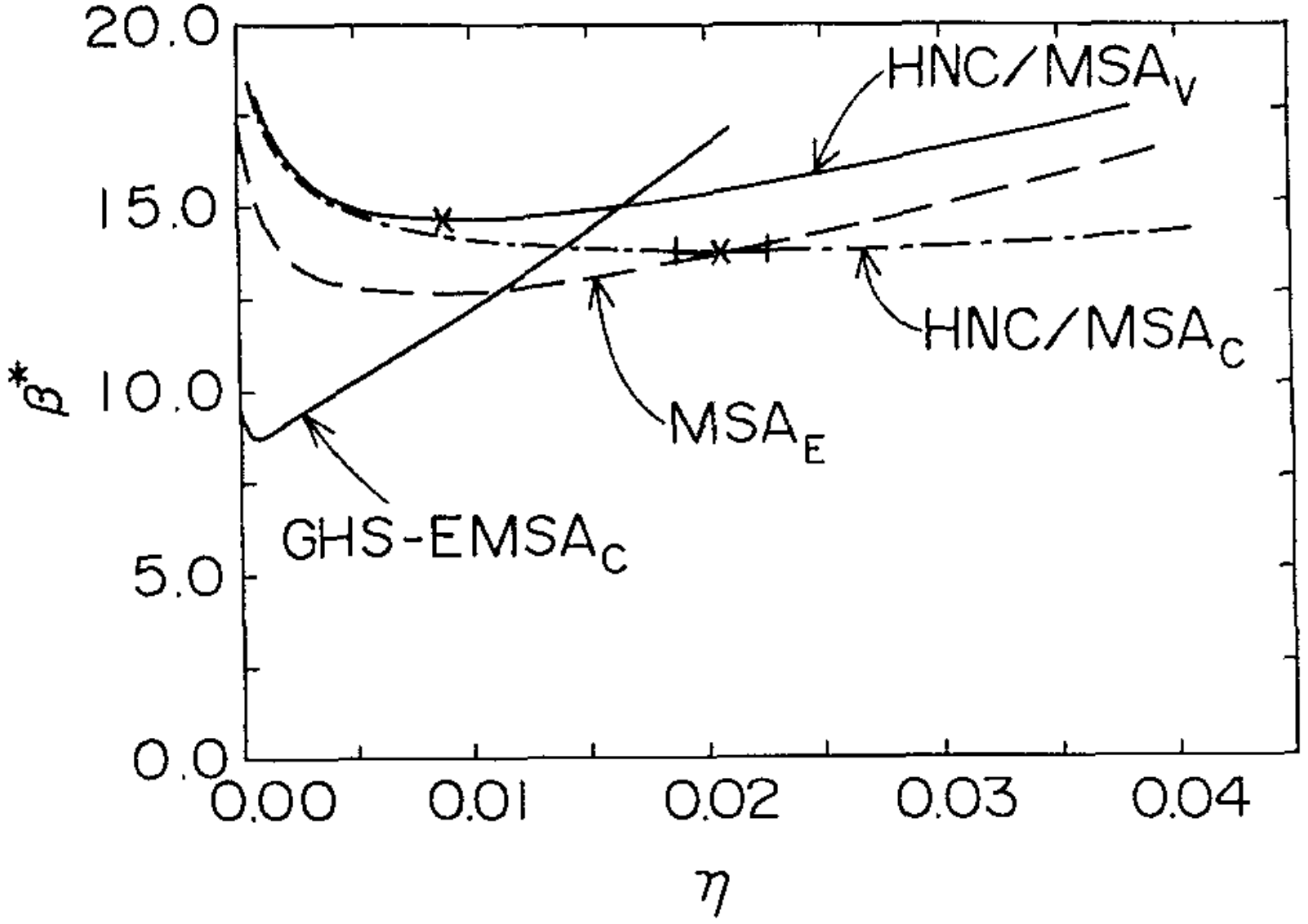}
        \caption{Phase diagram of a z:z electrolyte like NaCl where $n$ is the cation concentration.
                      Lines show the position of the spinodal using methods appropriate for each theory, and the minimum
                      indicates a critical point for fluctuation in ionic concentration.
                     Note the temperature axis is reversed by $\beta = z^2/d k_BT$ and $\eta = \pi n d^3/6$,
                     $d$ is the ion diameter.  Reprinted from \href{https://doi.org/10.1063/1.461516}{Ref.~\citenum{gtova91}},
                     with the permission of AIP publishing.}
        \label{f:KT}
    \end{subfigure}
 \caption{Comparing phase diagrams of the electron gas dissolved ions.
Both show an insulating phase at low density (labeled Wigner crystal in (a)) and
a conducting phase at high density separated by a minimum.
The corresponding transition in an electron gas has not been well studied, but critical temperatures
feature in the phase diagram of superconducting cuprates (where $n$ is percent of solid impurities).\cite{rlaug00}}\label{f:phase}
\end{figure}

  It turns out that the electron gas in `real' jellium behaves rather differently at low and
high density.  At low density, the electron positions are dominated by pairwise
repulsion, and organize themselves into a lattice (of plane waves) with low conductivity.\cite{pnoze58}
This low-density state is named the `Wigner lattice' after E. P. Wigner, who
computed energetics of an electron distribution based on the lattice symmetry of its host metal.\cite{ewign34}
At higher densities, collective motions of electrons screen out the pairwise
repulsion at long range.  This gives rise to a nearly `free,' continuous distribution
of electrons with higher conductivity more like we would picture for a metal.
Fig.~\ref{f:ceperly}, from a well-known particle-based simulation of Ceperly and Alder,\cite{dcepe80}
shows the Wigner lattice as well as both spin-polarized and unpolarized high-density states.

% The birth of screened Coulomb operators: range splitting
  Taking the opposing side, early applications of self-consistent field (Hartree-Fock or HF) theory
to molecules and oxides noticed that the long-range, collective `correlated' behavior of the electrons
was usually irrelevant to the short-range structure of electronic orbitals.
Getting the short-range orbital structures right allowed HF theory to do well describing
the shapes of molecules and the cohesive energy of metal oxides,\cite{nmott49} as well as
magnetic properties.\cite{ewign38}
More recent work has shown explicitly that a model that altogether omits the long-range
tail of the $1/r$ potential still allows accurate calculations of the lattice energy of salt crystals.\cite{pgill96}

  Although both theories worked well for their respective problems,
the transition from insulating to conducting metals (as electron density increases)
also proved to be difficult because it involved a cross-over between both short- and long-range effects.
Because of this mixture of size scales required, relying exclusively on a theory appropriate
for either short- or long-range produces results that increasingly depend on
cancellation of errors.  This sort of error cancellation is illustrated by the
phenomenology of `overdelocalization.' 

  Well known to density functional theorists,
`overdelocalization' is the tendency of continuum  models for electron densities
(having their roots in the long-range TF theory)
to spread electrons out too far away from the nucleus of atoms.
The result is that electron clouds appear `softer' in these theories, and polarization
of the charge cloud by the charge density of a far molecules contributes too much energy.
On the other hand, induced-dipole induced-dipole dispersion forces are not modeled by
simple density functionals, and so their stabilizing effect is not present.
It has been found that the over-delocalization can be fixed by making a physical distinction
between short and long-range forces.  However, the resulting binding energies are not strong enough.
After the correction, they need a separate addition of a dispersion energy to bring them
back into agreement with more accurate calculations.\cite{msoni15}  Thus, a bit of sloppiness
on modeling short-range structure can compensate for the missing, collective long-range effects.
% The behavior of the SR model.
% Trouble combining SR with LR - corrections disrupt `Pauling point' of density overdelocalization compensating dispersion (many-body correlations)

\section*{Hybrid Theories in Electronic Structure}  
% Born-Oppenheimer Approx. - frozen nucleii with conti. electronic densities

  When looking at properties like the cross-over between
conducting and insulating behavior of electrons, it's not surprising that
successful theories strike a balance between short-range, discrete structure
and long-range continuum effects.  Even in the venerable Born-Oppenheimer
approximation from 1927, we see that atomic nucleii are treated as atoms (immovable point charges),
while electrons are described using the wave theory.  The separation in time-scales
of their motion makes this work.  By the time the atoms in a molecule
have even slightly moved, the electrons have zipped back and forth between them
many times over.

% explain characterization of SR/LR in terms of the structure factor and RDF
  Correlation functions are a central physical concept in the debate between
long and short range ideas.  The distance-dependent correlation function, $g(r)$,
measures the relative likelihood of finding an electron at the point,
$r$, given that one sits at the origin.
One of the first attempts at accounting for electron-electron interaction
was to use perturbation theory to add electron
interactions back into the uniform gas model ($g(r) = 1$).
The first order perturbation modifies this by looking at interactions
between electrons of the same spin.
This interaction is termed the exchange energy,
since it comes from pairs of electrons with
the same spin exchanging momentum.\cite{pines3}
After the correction, electrons with parallel spin now have
smaller density at contact, $g(r) = 1-\tfrac{9}{2}(\sin (k_F r) - k_F r \cos(k_F r))^2/(k_F r)^6$.

% introduce particle simulations, MC and QMC
  The correlation function between infinite periodic structures
is, $S(k)$, the long-range analogue of $g(r)$ (in fact its Fourier transform).
The function $S(k)$ is called the structure factor by crystallographers.
If the system consisted only of electrons, the structure factor could be measured
directly by light or electron scattering experiments.
There, $S(k)$ is the intensity scattered out at angle $\theta = 2\arcsin(\lambda k/4\pi)$
when the material is placed into a weak beam of photons or electrons
of wavelength $\lambda$ pointed in the $\theta = 0$ direction.
This function has been computed using an accurate particle simulation technique
and shown in Fig.~\ref{f:ueg}e,f.\cite{gorti94}
The curves are labeled by $r_s = (3/4\pi n)^{1/3}$, measured
in units of Bohr radii.

  There is a duality between short and long range perspectives
inherent in $g(r)$ and $S(k)$ as well.
Long-range behavior appears at large $r$ when $g(r)$ approaches 1.
At small $r$, the geometry of inter-particle interactions
determines the shape of $g(r)$.
Because particle dynamics is carried out in real-space,
$g(r)$ tends to be used by its practitioners to characterize
short and long-range structure.
Analytical solutions of many models, and especially those aiding
experimental measurements, are simpler
in Fourier space.  There, $S(0)$ is the integral of $g(r)$.  It provides
information on the total fluctuations in the number of particles,
and is a long-range quantity from which the compressibility,
partial molar volumes, and other properties can be computed.\cite{pdebe87}
Short-range structures that repeat with length $d$ show up
as peaks in $S(k)$ at correspondingly large $k=2\pi/d$.

% Bohm-Pines collective variable approach - TODO: add Bohm/Pines cites.
  Back to the metallic/insulator problem, between 1950 and 1953 Bohm and Pines pioneered the idea of
explicitly splitting the energy function (Hamiltonian) governing electron motion into
local and long-range degrees of freedom.\cite{dbohm50,bohm,phugh06}  Using the intuition that long-range
collective motions of electrons should look like the continuous plane-wave solutions to Maxwell's theory,
they added and subtracted those terms and called them `plasmons.' (Fig.~\ref{f:hybrid}d)
Just like photons, the plasmons are continuous waves when treated classically, but are
quantized particles when understood quantum mechanically.

  What remained after the subtraction was a Hamiltonian whose interactions were
only short-ranged, but could not be treated with a continuum description.
Instead, the short-range part describes interactions between effective discrete particles
which Bohm and Pines dubbed `quasiparticles.'
The quasiparticles were like packs of electrons surrounded
by empty space, `holes.' The quasiparticles thus
have larger mass and softer, screened, pair interactions (explaining
why the mass has to be fixed when applying the free electron theory to metals).
These new `renormalized' electron quasiparticles could even
have effective pairwise attraction.  This latter effect
was a central component to the BCS model of superconductivity,
where the quasiparticles are known as `Cooper pairs.'
Because of its dual representation, the Bohm-Pines model gave good
answers for both cohesive energies and conductivities -- and described the cross-over
between insulating and metallic regimes as electron density is increased.\cite{pnoze58}

% Density functional methods - exact with "unknown" right answer
  For all its descriptive power, the Bohm-Pines approach was often lamented for its
requirement for a specific set of approximations.  Most damningly, it required inventing a continuum
of plasmons to describe the long-range interactions of a finite set of electrons.
This adds infinite degrees of freedom to a system with an initially finite number.
It also required the plasmons to stop and the particles to commence at some cutoff wavelength.
These troubles lead us into the problem of renormalization group theory,
which is beyond the scope of the present article.

  In fact, in 1954, just after the publication of the last article in the Bohm and Pines
series above, Lindhard provided a model for
collective electronic response of a metal that involved only the % TODO - cite
metal's correlation function (by means of its dielectric coefficient, $\epsilon$).\cite{phugh06}
Following a decade later in 1964-65 was Hohenberg, Kohn and Sham's
density functional theory.\cite{phohe64,wkohn65,phohe90}  Both developments
rephrased the description of electronic structure in terms of a continuous
field of electron density.  Linear response (perturbation) theory says
that an initially homogeneous density $n_0$ responds to an applied field, $\phi$ as,
\begin{equation}
\Delta n(r) = n_0 \int \chi(r,r') \phi(r'), \label{e:pert}
\end{equation}
where $\chi(r,r')$ is the Fourier transform of the structure factor above.
Their defining characteristic is the focus on continuous
response of that density to a continuous external field, $\rho(r) = \rho[\phi(r')](r)$.

% Step 2: describe the dielectric and RPA as related to g(r) and S(r)
  The theory may be understood as a fully long-ranged point of view
that includes short-range effects indirectly through $S(k)$.
It shows how to use integration to calculate all thermodynamic quantities
from structure factor.  The only problem is that it does not broach the issue
of how to predict the structure factor.  One well-known
method is to assume the probability of $n(r)$ is a Gaussian
on function space (so the exponent depends on $\int n(k)^2/\chi(k) \, dk^3$,
and $\chi(k)$ is just slightly different from $S(k)$).
In that case, the inverse of the correlation function ($1/\chi(k)$) is a self-energy term plus the inter-particle energy function.
This assumption is known as the random phase approximation (RPA),
named because of its historical discovery by Bohm and
Pines following from neglecting couplings between a set of linearly independent
(Fourier) modes, $n(k)$.
This ends up excluding all non-Gaussian fluctuations.

% TODO: describe basis of DFT formalism in LR fields
  The `dielectric' ideas encapsulated in the linear response theory of Eq.~\ref{e:pert}
can be combined with the free electron model of Eq.~\ref{e:TF} ($T[n]$ proportional to $n^{5/3}$),
or a wavefunction calculation of the kinetic energy, $T[n]$,
to synthesize modern density functional theory (DFT).\cite{eschrig,nlang70}
It writes the electron configuration energy as,
\begin{equation}
A[\phi] = \inf_{n(r)} T[n] + E_\text{XC}[n] + \int
    n(r) \left( \phi(r) + \frac{1}{2} \int dr'^3 \frac{n(r')}{4\pi\epsilon_0 |r-r'|} \right)\, dr^3
.
\end{equation}
Now the (long-range) correlation function of the electron, $\chi$, is obtained from the curvature
of $A[\phi]$.  Mathematically, the unknown structure factor has been migrated into an unknown
functional, $E_\text{XC}[n]$.  The initials stand for exchange and correlation, its two major
components.  The principle advantage gained by this rephrasing
is that new, accurately known (usually short-range) terms like $T[n]$ can be added to
$A[\phi]$ in order to decrease the burden on $E_\text{XC}$ to model `everything else.'
The disconnect between short and long-range energies can be shoveled into some fitting parameters.

  Again moving forward 40 years, the relative unimportance of long-range Coulomb
interactions for local structuring noticed by Lang and Perdew\cite{dlang77,pgill96} lead to the
suggestion that the density functional method itself should also distinguish between
short and long range structural effects.  Implementation of this idea was perhaps
first carried out by Toulouse, Colonna and Savin in 2004.\cite{jtolo04}
There, the local density approximation deriving its roots in the TF theory
is applied to describe short-range interactions, while the HF theory
is used to ensure proper electron-pair repulsion (exchange) energies at long-range.
The association of HF with long-range and density functional (DF) with short-range
apparently runs counter to our association between continuum, density-based,
models for long-range interactions {\em vs.} discrete, particle-based models for short-range interactions.
A major complication with our association is that it is known that the HF method
describes the long-range (asymptotic) electronic interactions well, whereas the DF method does not.
DF methods were historically used to describe the `entire' energy
function, and have thus been tailored to describe quasi-particles (the so-called exchange hole),
rather than asymptotics.
This association was put to the test shortly after by Vydrov and co.\cite{ovydr06}
using an earlier DF called LSDA that is not strongly tailored in this way.
They separately averaged the short- and long-range components of HF and DF
and checked their ability to predict the cohesive, formation energies of small molecules.
Doing so, they discovered that models with no HF at long range had similar
descriptive power to those that used only DF
at short range and only HF at long range.
Split-range functionals are still an evolving research topic.

\section*{Liquid-State Theories}

% Diffusion (from Jaynes)
% Ehrenfest / Jaynes debate
% Boltzmann transport equation (from Jaynes)
  The divide between short and long-range, discrete, and continuous distributions
also plays a key role in the development of thermodynamic theories for gasses and liquids.
In the 1860s, Boltzmann proposed his transport equation for the motion of gas density
over space and time.  The model employed the famous sto\ss{}zahlansatz, which states that the
initial positions of molecules {\em before} each collision is chosen `at random.' (Fig.~\ref{f:hybrid}a)
In the original theory, the probability distribution over such random positions
was often confused with their statistical averages\cite{pehre12}
-- a point which lead to enormous confusion and controversy persisting
even until 1960.\cite{ejayn65}

  This history very nearly parallels the development of electronic density
theories.  After electromagnetism and gas dynamics had been worked out
at the end of the 19th century, Gibbs' treatise on statistical mechanics
laid out the classical foundations of the relationship between statistics and dynamics of molecular systems.
Nevertheless, there were contemporary arguments with Ehrenfest and others
about the need for introducing statistical hypotheses into an exact dynamical
theory.\cite{pehre59}  Early on, it had been hoped that an exact study of the motion of the
molecules themselves could predict the appropriate `statistical ensemble'
by finding long-time limiting distributions.
However, that hope was spoiled by the notice that initial conditions must be described statistically. %\cite{}
The idea persists even at present, though it has been tempered by the recognition
that sustaining nonequilibrium situations requires an infinitely extended environment,
which has to be represented in an essentially statistical way.\cite{ggall16}

  The resolution, according to Jaynes,\cite{ejayn79} is to understand the Boltzmann transport
equation as governing the 1-particle probability distribution, $N P(r|C)$, rather than the
average amount of mass, $n(r)$, at point $r$.  It turns out that this switch in perspective from exact knowledge
of all particle positions to probability distributions is one of the key ways
of separating short and long-range effects.
% 1923 - Debye model
Two of the oldest and most widely known uses of this method are in the
dielectric continuum theory dating from before Maxwell's 1870 treatise,
even to Sommerfeld (Fig.~\ref{f:hybrid}c), %\cite{}
and the Debye model of ionic screening from 1923.
For both, a spatial field $E(r-r_0)$, emanating from a discrete molecule at $r_0$,
is put to a bulk thermodynamic system whose average properties
are well-defined using, for example, $P(r | E)$ for the dipole density $\mu(r)$ at point $r$, due to
a field, $E$ or $n(r; \phi)$ for the ion density at point $r$ due to a voltage, $\phi$.
Treating $\phi$ and $E$ as weak perturbations and looping $\mu(r)$ (or  $n(r)$)
back in as additional sources gives a self-consistent
equation for the response of a continuum.

% TODO: core DFT
  As was the case for electronic structure theory, the most concise description
of this type of self-consistent loop is provided by a density
functional equation for the Helmholtz free energy (with $\beta = 1/k_BT$),
\begin{wrapfigure}{r}{0.5\textwidth}
{\centering
\includegraphics[width=0.48\textwidth]{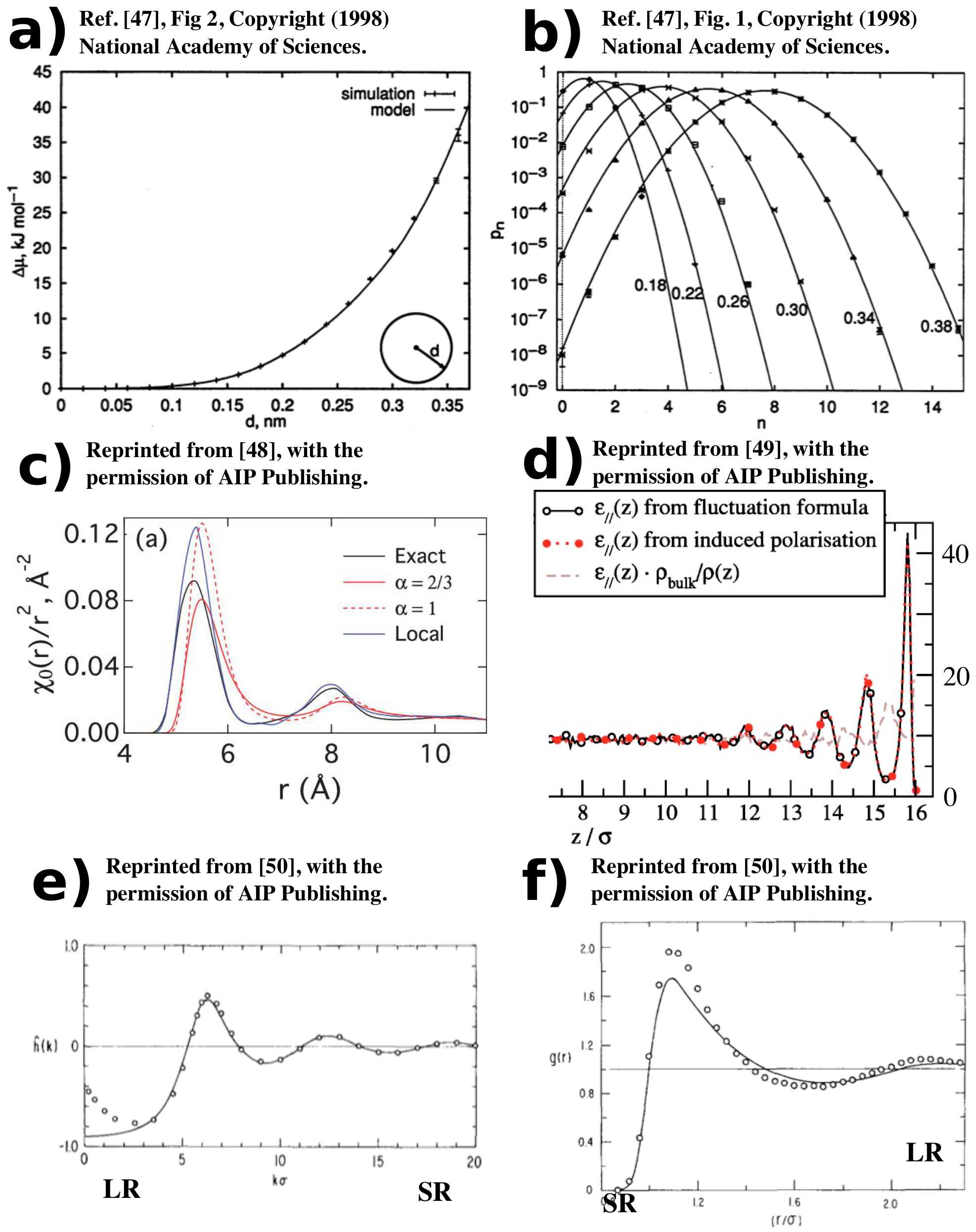}
\caption{Short- and long-range theories of solvent dipole and electrolyte structure.
(a) and (b) show free energies and number occupancy distribution for spherical
cavities in water.\cite{ghumm96}
(c) shows the dielectric response in a spherical geometry\cite{dmoha15}
and (d) shows the dielectric permittivity computed in a slab geometry.\cite{vball05}
(e) and (f) show the correlation function of a supercritical Lennard-Jones fluid
near $n = 0.52/\sigma^3$, $T = 1.34\epsilon/k_B$.\cite{wca}}\label{f:ions}
}\vspace{-4em}
\end{wrapfigure}
\begin{align}
\beta A[E, \beta] &= \inf_{\mu\in\{\mu\}} \left[ -\log( g[\mu] ) - \beta E\cdot \mu \right] \notag \\
  &\approx -\ln \sum_{\mu\in \{\mu\}} g(\mu) e^{\beta E \cdot \mu}
.
\end{align}
The curvature of $A$ with changing applied field, $E$, gives the response
function which is related to the conventional dielectric.
Consider first a case where $\mu$ contains enough information
 to exactly assign
a dipole to every one of $N$ molecules.  An example would be a single molecule
with twice as many ways to create a small dipole as a large one,
$g(4 \text{ D}) = 2$ and $g(2 \text{ D}) = 4$ (1D = 1 Debye).
Then $g(\mu)$ is a product over counting factors.
The free energy, $A$, will have jump discontinuities
in its slope as the field, $E$ is varied because the solution jumps from one assignment ($\mu=2$ D)
to another ($\mu=4$ D at $\beta E \ge (\ln 2)/(2 \text{ D})$).  Its graph is very much like Fig.~\ref{f:ueg}a.
In a discrete function space, density functional theory equations yield
solutions exhibiting the a discrete nature.

  On the other hand, if $g(\mu)$ varies continuously
with $\mu$ in some range of allowed average densities, then the solution
will describe a smooth field free energy.  Interestingly, starting from the first
situation and computing
\begin{equation}
S[\bar \mu, \beta] = -\beta \sup_E \left[ A[E,\beta] + \int E(r)\cdot \bar \mu(r) dr\right]
,
\end{equation}
leads to such a continuous version of $\log g(\mu) \approx S(\mu)$ (in fact its concave hull).
This concave function allows densities that are intermediate between
discrete possibilities for the system's state.
Such intermediate densities could only be reached physically
by averaging, so that $\bar \mu$ is an average polarization over
possible absolute assignments of dipoles to molecules, $\mu$.

  After the theory of self-consistent response to a long-range field had been worked out,
further development of liquid-state theory had to wait 40 years for developments
in quantum-mechanical interpretation of light absorption
and scattering experiments.  Some early history is given in Ref.~\citenum{hkrag18}
and Debye's 1936 lecture\cite{pdeby36} in which he explains how electronic
and dipole orientational polarization could be clearly distinguished
from measurements of the dielectric capacitance of gasses
along with the great advancements made in the 1920s (which Debye credits
to von Lau in 1912) of using x-ray and electron scattering to confirm molecular structures already
adduced by chemists from symmetry and chemical formulas alone.
Thus, the long-range theory gave a comprehensive enough description of macroscopic
electrical and density response that it could be used as a basis
to experimentally determine local structure.

  With statistical mechanics, quantum mechanics, and molecular structure
in hand, liquid-state theories developed in the 1930s-50s through testing
hypotheses about the partition function against experimental results
for heat capacities.  One of the earliest models was the `free volume'
(also known as cell model) theory, developed by Eyring and colleagues and independently
by Lennard-Jones and Devonshire in 1937.  The theory was put
on a statistical mechanical basis by Kirkwood in 1950,\cite{jkirk50}
as essentially expressing the free energy of a fluid in terms of
the free energy of a solid composed of freely moving molecules
trapped, one each, in cages exactly the size of the molecular volume, plus the free
energy cost for trapping all the molecules
in those cages in the first place.
It competed\cite{dmcqu62} with the `significant structure' theory of liquids
(also proffered by Eyring and colleagues\cite{heyri58,heyri63}).
In the significant structure theory (Fig.~\ref{f:hybrid}f), the partition function for the fluid
is described as an average of gas-like and solid-like partition functions
to account for the difference in properties between highly ordered and more
disordered regions (which contain vacancies).

% The HS fluid and PY/MSA
\subsection*{ Scaled Particle {\em vs} Integral Equations}
% scaled particle theory: (Riess) -- bridge SR E ~ rho V and LR surface tension
  Also around that time, a competition emerged between the
scaled particle theory\cite{hreis59} and the `integral equation'
approach based on (and now lumped together with)
Percus and Yevick's\cite{jperc58,jperc62} closure of a theory created by
Ornstein and Zernike in 1914 to calculate the effect of correlated
density fluctuations on the intensity of light scattered by
critically opalescent fluids.\cite{lorns14}
% TODO - cite OZ
This connection was significant, since theories
of the correlation function prior to 1958 applied the
superposition approximation due to Kirkwood, Yvon, Born, and Green (ca. 1935).\cite{davis,abenn06}

\begin{wrapfigure}{r}{0.5\textwidth}
{\centering
\includegraphics[width=0.45\textwidth]{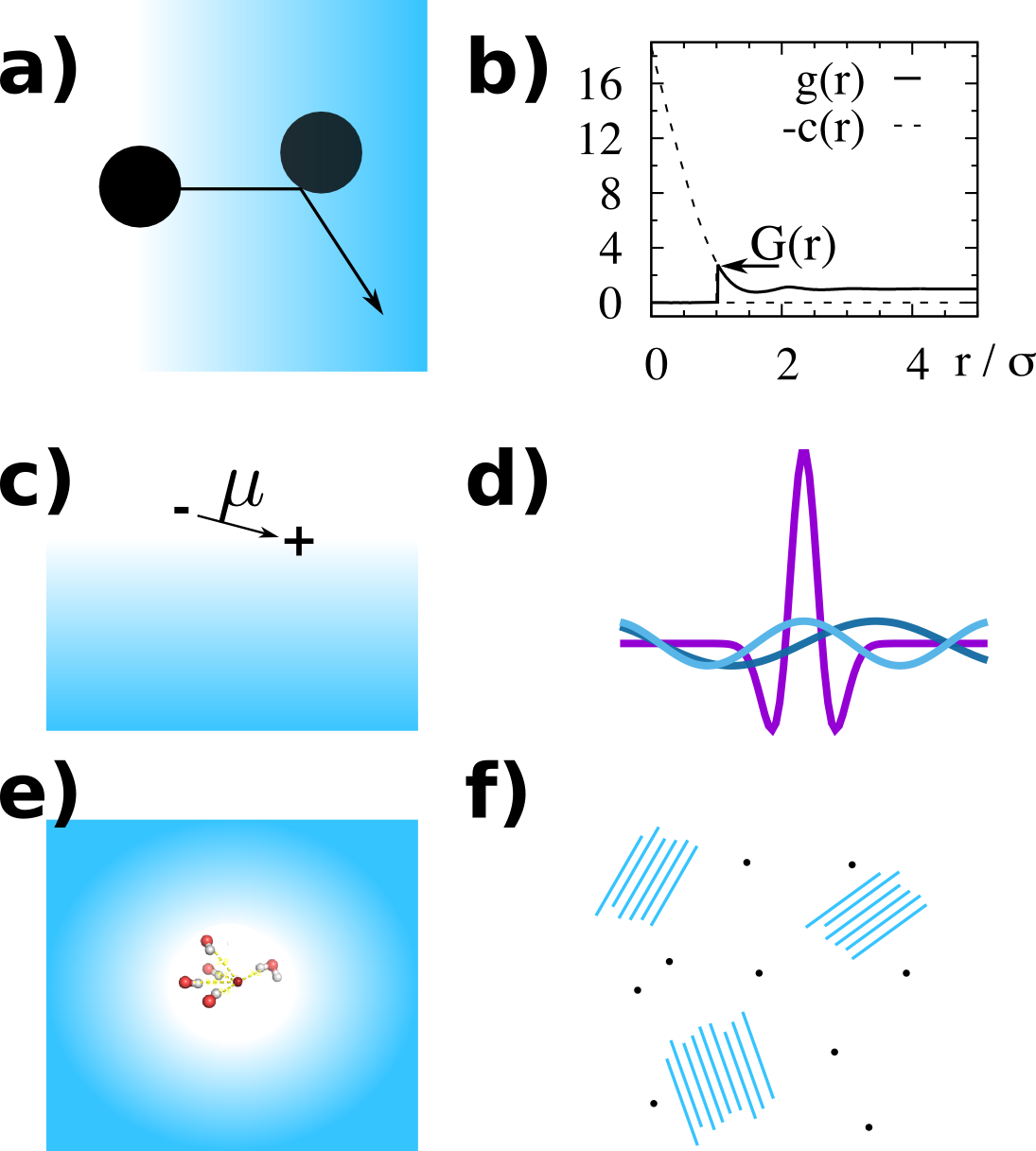}
\caption{Hybrid discrete/continuum theories. (a) Boltzmann picture of scattering by one particle chosen `at random' from the
continuum. (b) Mean spherical approximation for the hard sphere fluid of diameter $\sigma$.  $g(r)$ and $c(r)$ are known at $r<<\sigma$ and $r >> \sigma$, but the central region is a guess.  (c) Sommerfield conception of a dipole above a continuous polarizable medium. (d) Bohm-Pines conception of a quasiparticle (purple, central peak) and two long-range plasmons (blue).  (e) Dressed ion, quasichemical, or Lorenz-Lorentz-Mossotti-Clausius\cite{hkrag18} cavity models of a discrete molecule in a continuum solvent, (f) significant / inherent structure theory of a coexisting mixture of ordered and disordered regions making up an overall homogeneous phase.}\label{f:hybrid}
}\vspace{-3em}
\end{wrapfigure}

% Explain the nature of both theories
  The scaled particle theory (SPT) approach takes the viewpoint that
the number, sizes and shapes of molecules in a fluid are determined
by integrating the work of `growing'  a new solute particle in the middle of a fluid.
Its organizing idea is that the chemical potential of a hydrophobic solute
is equal to the work of forming a nanobubble in solvent.
%\begin{equation}
%\pd{\beta \mu}{\lambda} = \avg{ e^{\beta\Delta U_\lambda} \pd{}{\lambda}e^{-\beta\Delta U_\lambda}  }_\lambda
%.
%\end{equation}
%Here, $\lambda$ is a parameter determining the shape of the solute molecule
%and $\Delta U_\lambda$ is the interaction energy of the solvent with a solute
%of shape $\lambda$.
For simple hard spheres, the work is $P dV$, where $P = k_B T n_0 G(d)$,
$n_0$ is the bulk solvent density, and $G(d)$ (Fig.~\ref{f:hybrid}b),
the density of solvent molecules on the surface of
the solute of diameter $d$.
Hence, knowing the contact density for any shape of solute molecule
provides complete information on the chemical potentials of those molecules.
This very local idea can be related to counting principles at very small sizes,\cite{hashb06}
and continued through to macroscopic ideas about surface tension at very large sizes -- creating
a way to interpolate between the two scales.

  On the other hand, the integral equation approach expresses the idea that long-range
fluctuations in density are well described by a multivariate Gaussian
distribution.  If the probability distribution of the density, $n(r)$,
was actually Gaussian, its probability would be,\cite{dcall96}
\begin{equation}
P[n(r)] = P[n_0] \exp\left( -\frac{\beta}{2} \iint dr dr' \, (n(r)-n_0) G(r,r') (n(r')-n_0) \right) / Z[\beta G],
\label{e:Pn}
\end{equation}
where $G(r,r') \equiv \mathrm{const}\cdot\delta(r-r') - c(r-r')/\beta$.
In the RPA, $-c(r)/\beta$ is energy for placing a pair of molecules at positions $r$ and $r'$.\cite{dfryd16}

  When they are not Gaussian distributed, the correlations in instantaneous
densities, $n(r)$, provide a means of estimating $c$, the direct correlation function.\cite{knewm94}
This long-range idea has been used to show that $G$ degenerates to
the pairwise energy for very large separations ($G(r) \to U(r)$ as $r\to\infty$).
For simple hard spheres, it can also be related to counting principles at short separations, since there
the correlations must drop to -1, expressing perfect exclusion.
Assuming limits both hold right up to the discrete boundary
of a solute yields the mean spherical approximation (MSA, Fig.~\ref{f:hybrid}b).

  These two theories thus express, in pure form, the
divide between short-range and long-range viewpoints on molecular structure.
Integral equation theories are most correct for describing continuum densities
and smooth interactions.  Theories that, like SPT, are based on
occupancy probabilities of particles in well-defined local structures and
geometries are most correct for describing short-range interactions that
can contain large energies and discontinuous jumps.

  Fig.~\ref{f:ions}b shows $P(n|d)$, the probability that a
randomly chosen sphere of radius $d$ contains exactly
$n$ discrete water molecules.  Each curve is marked by its value
of $d$ in nanometers.  The free energy for creating an empty
nanobubble of size $d$ in water is shown in its counterpart,
Fig.~\ref{f:ions}a.  Both computations are very closely
related, and easiest to do from the local picture of scaled particle theory.
The cavity formation free
energy (Fig.~\ref{f:ions}a) is, in principle, also able
to be computed from a density functional based on relating
the logarithm of Eq.~\ref{e:Pn} with the entropy.\cite{dcall96}
However, when the calculation is done in the usual density functional
way the cavity formation free energy is surprisingly difficult to reproduce.\cite{gjean13,gjean15}
This difficulty is related to the abrupt decrease in solvent
density to zero at the cavity surface.  In addition to mathematical
difficulties,\cite{jchay84a} this complicates creating
a physically consistent functional from bulk properties alone.
From scaled particle theory, we know the
free energy should scale with the logarithm of the volume for small
cavities, but later switch over to scale with the surface area.
The transition distance is determined by the size of discrete solvent molecules.

\subsection*{ Perturbation Theories}

  Slowly but surely during the same time period as integral equation theories
were being developed the method of molecular dynamics
emerged.\cite{lverl67}  Its primary limitations of small, fixed, particle numbers,
large numbers of parameters, finite sizes and short timescale simulations
weigh heavy on the minds of its practitioners.\cite{jjoh93}  Early models of water
needed several iterations before reproducing densities,
vaporization enthalpies and radial distribution functions from experiment.
Initial radial distributions from experiment were wrong, and the models
had to be corrected and then un-corrected to chase after them.\cite{asope13}
Surprisingly, early calculations took the time and effort to calculate
scattering functions and frequency-dependent dielectrics to compare to experiment.\cite{arahm71,fstil82,hfried73}
By contrast, the bulk of `modern' simulations report only the data that can be readily
calculated without building new software.

  By checking data from integral equations against molecular dynamics (MD)
and scattering experiments it was clear by 1976 that
many powerful and predictive methods had been created to describe
the theory of liquids.\cite{jbark76,hsing83}
Nevertheless, there remained even then lingering questions about the
applicability of integral methods to fluids where molecules contained dipole moments,
and the treatment of long-range electrostatics in MD.
Some difficulties in modeling phase transitions and interfaces
were anticipated, but it was hardly expected that bulk molecular dynamics methods
themselves would stall and eventually break down when simulating liquid/vapor and liquid/solid surfaces.

  This trouble is illustrated by the simulation community's reception of the work
leading to Fig.~\ref{f:ions}c,b.  Both show the dielectric response function for water
dipoles at the interface with a large spherical particle (left) or vacuum (right).
The latter shows a correlation function computed from all-atom molecular dynamics
by Ballenegger.\cite{vball05}  This full computation was preceded two years earlier
by less well-cited theoretical work from the same author.\cite{vball03}
As of writing, the citations counts are 140 and 19, respectively.
Even after its publication, the technical difficulties
caused by simulating collective dipole correlations
inside a finite size box cast a cloud over the interpretation that drove Ballenegger
back into those fine details for the following nine years.\cite{vball09,vball14}
On the left (Fig.~\ref{f:ions}c) is a simulation of water's dipolar response
next to a large sphere.\cite{dmoha15}  The finite-size effects are less severe,
and a comparison (not common in contemporary literature) is made to analytical
theories that apply to infinite systems.  However, those analytical theories work
best at long-range, and disagree on the short-range
order.  The disagreement is jarring because energetic contributions of long
and short-range order are on the same order of magnitude.

% The LJ fluid spoils all that - virial vs. compressibility routes
% Compressibility for gasses and plasmas (important LR contributions)
% Virial for HS and well-structured fluids (inherent structures again)
  It was also beginning to be recognized that there were two complementary
approaches to the theory of fluid structure.  The short-range viewpoint
stated that the radial distribution function should be reproduced well
at small intermolecular separations (small distance in real-space as in Fig.~\ref{f:ions}f).
This leads to good agreement with interaction energies and pressures
so that the virial and energy routes to the equation of state work well.\cite{wca}
The long-range viewpoint instead emphasizes reproducing the
structure factor at small wavevectors (as in Fig.~\ref{f:ions}e).
Because of this, it favors using the compressibility route
to the equation of state and leads to good agreement with fluctuation quantities.\cite{rperr88}

%  Water proved to be a major challenge to molecular models
%because of its mixture of short-range hydrogen bonding and long-range
%dipole order.  The first molecular forcefield models for water (focused
%on obtaining tetrahedral structure) did not yield the known enthalpy of
%vaporization or density.\cite{}  Later models were `tuned' to provide these
%values, but also needed to account for the fact that water molecules
%open up to a wider angle (lowering their dipole) upon escape into the gas phase.\cite{spc}

% Explain advent of perturbative theories
%The classical atomic model provided geometric detail on liquids
%It was then found that using a hard-sphere condition plus a classical,
%Coulomb electrostatic interaction was sufficient to predict the structure
%of most molecules.\cite{}

% Stillinger-Weber: inherent structures
\subsection*{ Inherent structures}

  Water proved to be a major challenge to molecular models
because of its mixture of short-range hydrogen bonding and long-range
dipole order.\cite{jmu76}  One successful physical picture of water was provided by the Stillinger-Weber
`inherent structure' model introduced in the early 1980s.\cite{fstil83}
It represented a cross between the `significant structure' 
theory and the free volume theory.  In it, molecules are fixed
to volumes defined by their energetic basins, rather than
by a rigid crystal lattice.  Where the free volume theory had only
one reference structure, the inherent structure (like the significant structure theory)
had many.  One for each basin.  Each energetic basin looks, on an intermediate
scale, like a distortion of one of the crystalline phases of ice.
Thermodynamic quantities can be predicted using
the energies and entropies
associated to each basin
-- by virtue of the minimum energy structure and the number
of thermal configurations mapping to that minimum.

\section*{ Hybrid Theories in Liquid-State Structure}

  The Lennard-Jones fluid presented a challenge to the integral equation
and scaled particle theories above because it contains both short-range repulsion
and long-range attraction.  At high densities, however, it was found that
the radial distribution function was almost identical to the radial distribution
for hard spheres (compare Fig.~\ref{f:ions}e and Fig.~\ref{f:hybrid}b).
The transition from liquid to solid was also described
fairly well using the hard-sphere model.
On the other hand, at low densities the distribution function could be
described by perturbation from the ideal gas.
These two discoveries justify
the use of a perturbation theory to calculate the effect of long-range interactions
at very low and very high densities.\cite{lverl72}
A comparison of molecular dynamics with integral equation plus correction
theories is shown in Figs.~\ref{f:ions}e,f.\cite{wca}

  At intermediate densities, however, a liquid-to-gas phase transition occurs
that can be qualitatively understood, but not explained well as a
perturbation from either limit.  Instead, the integral equation method
turns out to hold the best answer in the supercritical region.\cite{ccacc96}
It is often encountered in the form of a perturbation theory from the critical point.\cite{lreat99}
It is no accident that the integral equation method works well here.
Supercritical fluids are characterized by long-range correlations
that can take maximum advantage of that theory.
For the same reason, integral equations describe the compressibility well, but
do poorly on the intermolecular energy.

  Comparing to developments in electronic structure raises the
question of whether perturbation theory could fix the short-range
correlations in high and low density fluids.  This approach was popularized
by Widom's potential distribution theory.\cite{bwido82}
Its central idea is to drop a spherical void into a continuum of solvent,
and then to drop a solute into its center.
This divides the new molecule's chemical potential into a structural
part (due to cavity formation) and a long-range part (due to response of solvent to the molecule).
Originally, the former were based on a local density approximation from the hard
sphere fluid and the latter from a pairwise term that amounted to a van der Waals theory.

  Around 1999, this basic idea had been combined with
older notions about working with clusters of molecules
to create a new `quasi-chemical' theory.\cite{lprat99}
It refined the simple process of creating an empty sphere devoid of solvent
into that of creating a locally well-defined
cluster of solvent molecules.
The free energy required for this process
is still local and structural, but now the entire cluster
of solute plus solvent can be regarded as one, local, chemical entity.
In order to work with molecules that have `loose' solvent
clusters, a third step was also added.  After pulling solvent
molecules into a local structure and adding the long-range interactions
between solute and solvent, the third step
releases the solvent cluster, liberating any energy
that might have been trapped by freezing them.\cite{droge11}

  The opposite of this short-range-first approach could be an inverse
perturbation theory -- first deciding on the long-range shape of correlation
functions and second correcting them for packing interactions
at short-range.  This kind of correction would look like an adjustment
to the solution of the Poisson-Boltzmann equation.
Such an approach may first have been presented in Refs.~\citenum{blee96,iboru97},
and followed with interesting modifications of the Debye theory.\cite{patta93,rkjel16,rkjel18}
%However, their treatment was a simple hard sphere-like correction
%to the low density Debye-H\"{u}ckel limit.
%
%  Counter to the historical development of exchange corrections to
%TF theory, there did not seem to be a general, local correction of the
%long-range continuum theory until recently,
Even more recently, the basic idea was rigorously applied to
molecular simulation models by Remsing and Weeks. %a new way of looking at their self-consistent molecular field theory.
Their scheme eliminates a hard step between short and long-range
in the first step by splitting the Coulomb pair potential into smooth, long-range and
sharp, short-range parts.  The long-range forces (from the smooth part of the potential)
are used to compute a `starting' density using RPA-like perturbation from a uniform fluid.
Although it seems a lot like the molecular density functional method,\cite{tsluc81,szhao11,abenn06}
the density after the first step remains smooth at the origin, lacking any hard edges.
It has previously been considered under the title
`ultrasoft restricted primitive model.'\cite{aniko12}
Remsing and Weeks added a final step to this model
to create a cavity at the origin and compared the results to MD simulations.

\begin{figure}
    \centering
    \begin{subfigure}[t]{0.48\textwidth}
        \includegraphics[width=\textwidth]{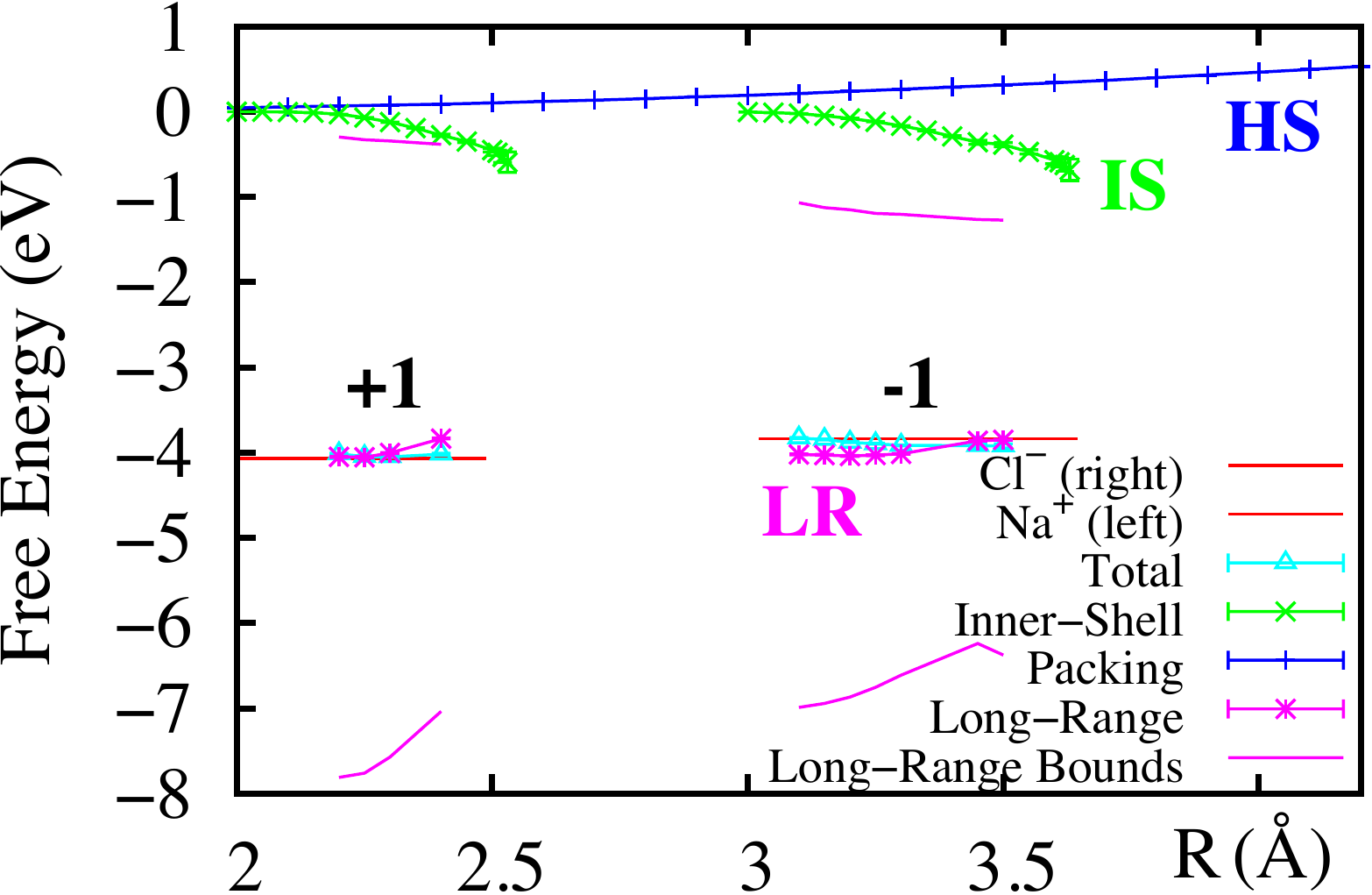}
        \caption{Ion solvation free energy components for the short-range (empty cavity first) model
computed from an MD model of NaCl in SPC/E water.  $R$ is the cavity radius,
`HS' denotes the cavity formation cost, `LR' is the full ion-SPC/E water interaction
after a cavity is present, and `IS' is the free energy of removing the cavity constraint.}
        \label{f:ionFE}
    \end{subfigure} \,
    \begin{subfigure}[t]{0.48\textwidth}
        \centering
        \includegraphics[width=0.8\textwidth]{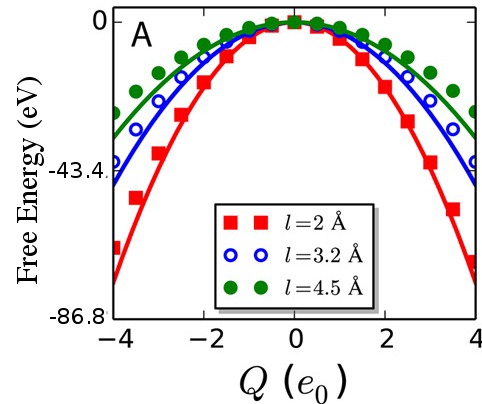}
        \caption{Interaction free energy of SPC/E water with a Gaussian charge distribution,
        		     $Q\exp(-r^2/l^2)/(l\sqrt{\pi})^3$.  Points correspond to simulation data, while lines assume
		     a constant dielectric model.
                     Adapted with permission from Ref.~\citenum{rrems16}, Copyright 2016 American Chemical Society.}
        \label{f:ionRR}
    \end{subfigure}
\caption{ Comparing components of the SR-first (left) and LR-first (right) calculations
of the free energy gained on dissolving a charged ionic species in water.}\label{f:fig5}
\vspace{-1em}
\end{figure}

%  Counter to the historical development of exchange corrections to
%TF theory, the method of adding long-range perturbation on top of
%known short-range energies appeared before the works above.
 
  Detailed molecular simulations have been used to compare the two approaches
with exact simulations by brute force calculation of all the energetic contributions.
Focusing on the short-range structure leads to a model whose first step is
to form an empty cavity in solution (blue curve in Fig.~\ref{f:ionFE}, labeled
`Packing').  Fig.~\ref{f:ionFE} shows the free energies of the next step
(Na$^+$ and Cl$^-$ ions) divided into `long-range' and `inner-shell' parts of the
re-structuring.\cite{droge08}  All points come from MD.
If, instead, the long-range interaction between an ion and solvent occurs first,
we are lead to couple the solvent to the smooth electric field of
a Gaussian charge distribution.  Fig.~\ref{f:ionRR} shows the free energy of that first
step as a function of charge for a variety of Gaussian (smoothing) widths.
The lines show continuum predictions, and the points show MD.
%These kind of join together when you look at the lower panel,
%where all possible steps I can think of were done.
%The application is really important, since it can pinpoint which free energy is
%most ``ion-specific.''

  Integral equation approaches to the dipolar solvation process have also
continued independently.
Matyushov developed a model for predicting the barrier to charge transfer reactions.\cite{nrft}
In that work, the dipole density response function to the electric field of a dipole
is worked out in linear approximation.  A sharp cutoff is used to set the
field to zero inside the solute, resulting in a hybrid short/long range theory.
The approach succeeds because the linear response approximation
(stating density changes are proportional to applied field)
is correct at long range, where the largest contributions to the solvation
energy of a dipole originate.
Other authors have expanded on numerical and practical aspects
of correlation functions.\cite{lding17,droge18,mvoro19}

% LMF theory from Weeks
%  Fast-forward two decades to find a difference of opinion on how to deal with
%(continuous) long-range structure of the solvent.  Weeks has presented a
%self-consistent field theory (LMF) that approximates the long-range intermolecular
%forces as continuous, and leaves the short-range ones as discrete.
%Pratt developed competing self-consistent field theory ideas.
%Recently, the two were happily reunited when considering an old application
%-- the pair correlation between two hydrophobes.
%This means we can try and cross-fertilize LMF and quasi-chemical theory ideas.

% Principle form for solvents: HS + LR + IS or LR + simulated local structuring?
  The theme of separating long-range, continuous {\em vs.} short-range, discrete
interactions runs throughout numerous other molecular-scale models.
Models in this category include the `dressed' ion theory, which posits that ions in solution always
go in clad with strongly bound, first shell, water molecules so that their radius is larger
than would be suggested from a perfect crystal (Fig.~\ref{f:hybrid}e).  These enlarged
radii appear in the Stokes-Einstein equation to describe the effect of molecular shape
on continuous water velocity fields when computing the diffusion coefficients for ions.\cite{skone98}
They should also appear to describe how excluded volume of ions will
affect the continuous charge distribution predicted by the primitive model
of electrolytes.  This modification is not common, and so would yield some nonstandard plots
of hydration free energy as a function of ion concentration.\cite{lblum91}
Solvent orientational order changes form again beyond about 1.5 micrometers
due to the finite speed of light.\cite{gkarl10}
The Marcus theory of electron transport describes two separate, localized
structural states of a charged molecule that interact with a continuously movable, long-range, Gaussian, field.
Larger magnitude fluctuations in the solvent structure lead to broader Gaussians, which in turn
are the cause of more frequent arrival at favorable conditions for the electron to jump.
It is common practice in quantum calculations to explicitly model all atoms and electrons
of a central molecule quantum-mechanically while representing the entirety of the
solvent with a continuous dielectric field.\cite{pren12,bmenn97,tduig13}

% Dressed ion theory - the hydrated radius concept for correcting Stokes-Einstein
% Marcus theory - continuous Gaussian solvent response around two principle structural states
% Implicit solvent quantum theory - continuous Gaussian solvent + conti. electron densities
% Issues with SR / LR overlaps - does the Pauli principle decrease el. polarizability in liquids?

  The theories above are not perfect.  They show issues precisely at the point where
short- and long-range forces are crossing over.  At high ionic concentrations,
the dressed ion theory breaks down due to competition between ion-water and ion-ion
pairing. %\cite{}
When solvent molecules are strongly bound, the use of a
continuous density field cannot fully capture their influence on thermodynamic properties.
Even without strongly bound solvent, dielectric solvation models leave open the
important question of whether electrons from the fully modeled molecule are more
or less likely to `spill out' into the surrounding solvent. %\cite{}
Returning back to Aristotle's objection to discrete objects, it is known that
density based models don't accurately capture the free energy of forming a empty cavity.\cite{gjean13,gjean15}
Thousands of years on, we are still vexed by the question of how
to understand the interface between material objects and vacuums.

\section*{The Future: A Middle Way}

  Early Eastern thought tends to place opposing ideas next to one another in an attempt
to understand them as parts of a whole picture.
Written around the beginning of the Middle ages, in 400 AD, the Lankavatara Sutra relates %Guatama
Buddha's view that this unity applies to atoms and `the elements' (which refer to something like the classical
Greek elements).
Taking liberties, we can say he is discussing a process
like instantaneous disappearance (annihilation) of a quantum particle in saying,
``even when closely examined until atoms are reached, it is
%not the destruction of the elements primary and secondary but of their 
[only the destruction of] external forms whereby
the elements assume different appearances as short or long;
but, in fact, nothing is destroyed in the elemental atoms.
What is seen as ceased to exist is the external formation of the elements.''
Bohr was well-known for his view on the `complementarity' principle,
stating in this context that the act of removing a particle makes its number more definite,
while making the amount of energy it exchanged with an external observer undefined.\cite{droge17}
Perhaps inspiring to Bohr sixteen centuries later,\cite{nbohr58}
the quote concludes, ``I am neither for permanency nor for impermanency ... there is no rising of the elements,
nor their disappearance, nor their continuation, nor their differentiation;
there are no such things as the elements primary and secondary;
because of discrimination there evolve the dualistic indications of perceived and perceiving;
when it is recognised that because of discrimination there is a duality,
the discussion concerning the existence and non-existence of the external world ceases
because Mind-only is understood.'' %\cite{lank}
%This sentiment is echoed around 605 AD by the Buddhist philosopher Seng-ts'an, credited with saying
%``The very small is as the very large when boundaries are forgotten;
%The very large is as the very small when its outlines are not seen.''
Bohr's complementarity could be contrasted with physicist John Wheeler.  He advocated, as a working hypothesis,
that participants elicit yes/no answers from the universe.  Replies come as discrete `bits,'
and are ultimately the reason that discrete structures emerge whenever
continuum models try to become precise.\cite{jwhee89}
Wheeler, in turn, could be contrasted with Hugh Everett, whose working
hypothesis was that the universe operates by pure wave mechanics.\cite{hever73,vlev18}
A modern resolution of those debates invokes small random, gravitational forces to
explain how quantum particles could become tied to definite locations.\cite{ldios18}
It is does not appear that there will be a resolution allowing us to do away with either
continuum or discrete notions.

%In the 1600s, just before Newton's time, Miyamoto Musashi wrote ``Students of the Nito Ichi School of
%strategy should train from the start with the [short] sword and the long sword in either hand...
%The long sword should be wielded broadly and the companion sword closely.''
  Of course, it is impossible to deduce scientific principles if we include any elements of
mysticism in a theory.  Nevertheless, the debate on the separation
between short and long-range seems to permeate history.
This idea that a meaningful understanding of collective phenomena should be sought
by combining physical models appropriate to atomic and macroscopic length scales
was taken up even recently by Laughlin, Pines, and co-workers.\cite{rlaug00}
They state, ``The search for the existence and universality of such rules,
the proof or disproof of organizing principles appropriate to the mesoscopic
domain, is called the middle way.''

  On one account it is clearly possible to set the record straight.
There are well-known ways of converting local structural theories
into macroscopic predictions and as vice-versa.
Bayes' theorem states that, for three pieces of information, $A$, $B$, and $C$,
\begin{equation}
P(A | B C) = \frac{P(B | A C) P(A|C)}{P(B|C)}
.
% P(A|C) = P(B|A) P(A|C) / P(B|C)
\end{equation}
If `C' represents a set of fixed conditions for an experiment,
`B' represents the outcome of a measurement,
and `A' represents a detailed description of the underlying physical mechanism
(for example complete atomic coordinates),
then Bayes' theorem explains how to assign a probability
to atomic coordinates for any given measurement, `B'.
Of course, in a reproducible experiment, $C$ will completely
determine $B$, so $B$ = $B(C)$.
Thus, the probability distribution over the coordinates
is a function only of the experimental conditions, $P(A|BC) = P(A|C)$.
This summarizes the process of assigning a local structural theory from
exactly reproducible experiments.

  On the other hand, a local structural theory provides an obvious
method for macroscopic prediction.  Given a complete description, `A,'
simply follow the laws of motion when interacting with a macroscopic measuring
device, `B.'   This would properly be expressed in the language
above as $P(B|AC) = P(B|A)$, since the experimental conditions are irrelevant.
Bayes' theorem then gives us a conundrum, $P(B|C) = P(B|A)$,
stating that every microscopic realization of an experiment must yield
an identical macroscopic outcome.

  The solution to the puzzle is to realize that unless an experiment is
exactly reproducible, $BC$ is always more informative
than the conditions, $C$, alone and $P(A|BC) \ne P(A|C)$.
This explains why studying exactly integrable dynamical systems is such a thorny issue,
and is the central conceptual hurdle passed when transitioning from classical to quantum mechanics.
Now identifying `B' with a partial measurement that provides a coarse scale observation
of some long-range properties, $P(A|BC)$ describes a distribution over
the short-range, atomistic, and discrete degrees of freedom.
Because of experimental uncertainty, the exact location of those
atoms is evidently subjective and unknowable (since it is based on measurement of $B$).
Nevertheless, it can in many cases be known to a high degree of accuracy.

% Bridging theories of linear and nonlinear response.
  Density functional theory traditionally focuses on $P(B|C)$, where `B' is the average
density of particles in a fluid and `C' is the experiment where a bulk material
is perturbed by placing an atom at the origin.  However, with a minor shift
in focus, $P(B|A' C)$ can also be found, representing the average density
under conditions where a particle is placed at the origin and some
atomic information, $A'$ is also known.  The objective of such a density functional
theory would be to more accurately know the long-range structure by including
some explicit information on the short-range structure.
The dual problem is to predict $P(A|B'C)$, the distribution over coordinates
when we are provided with some known information on the long-range structure.
In a complete generalization, we might focus instead on $P(AB|A'B'C)$,
representing the average density and particle distribution
under conditions where density and particle positions are known
only in part.  Bayes' theorem shows us that such a generalization
would just be the result of weaving the primal and dual problems together,
since (given the redundancies, $B' = B'(B)$ and $A' = A'(A)$),
$P(A | A'B'C) = P(A|B'C) / P(A' | B'C)$, and $P(B|A'B'C) = P(B|A'C) / P(B'|A'C)$.
%  It is not known whether such a theory would be helpful
%
%and the necessity of predicting $P(A'B'|C)$ in order to reconstruct the final
%result, $P(B|C)$.

% TODO: refer back to table of SR / LR correspondence
  The arguments above can be repeated for each of the elements in
Table~\ref{t:range} -- replacing SR with $A$ and LR with $B$.
What emerges is a persistent pattern of logical controversy,
where a problem can be apparently solved entirely from either
perspective.  In some areas, one of the other approach is
more expedient.  In every case, however, recognizing and using
both sides has proved to be profitable.
Comparing these two perspectives, we find that the discussion concerning
the existence of long and short-range theories ceases, leaving only
different ways to phrase probability distributions.

  We have now arrived at a point in the history of molecular science where these two great foundations,
short-range, discrete structures and long-range, continuum fields are at odds with one another.
Molecular dynamical models are fundamentally limited by the world view that all forces
must be computed from discrete particle locations.  Computational methods treating
continuum situations focus their attention on solving partial differential equations for situation-specific
boundary conditions.  Connecting the two, or even referring back to simple analytical models,
requires time and effort that is seen as scientifically unproductive.  Whats worse, it reminds us
that many, lucidly detailed, broad-ranging, and general answers were already presented
in the lengthy manuscripts which set forth those older, unfashionable models.

  Indeed, local and continuum theories are hardly on speaking terms.
In molecular dynamics, the mathematics of the Ewald method for using a Fourier-space sum
to compute long-range interactions are widely considered esoteric numerical details.
Much effort has been wasted debating different schemes for avoiding it by truncating and neglecting
the long-range terms.\cite{rwood95,hashb97,bmcca13}
On the positive side, the central issue of simulating charged particles
in an infinite hall of mirrors has been addressed by a few works.\cite{ssaka98,phune99,lbell17}
Much greater effort has been devoted to adding increasingly detailed parameters, such
as polarizability and advanced functional forms for conformation and dispersion energies,
to those atomic models.  Apparently, automating the parameterization process\cite{phuds18}
is unfundable.  In the case of polarization and dispersion, the goal of these
atomic parameters is, somewhat paradoxically, to more accurately model the long-range interactions.
The problem of coupling molecular simulations to stochastic radiation fields has, apparently,
never been considered as such.  Instead, we can find comparisons of numerical
time integration methods intended to enforce constant temperature on computed correlation functions.\cite{jbasc13}
In continuum models based on partial differential equations, actual molecular information
that should go into determining boundary conditions, like surface charge and slip length
(or, more accurately, boundary friction\cite{bcros18}),
are replaced by `fitting parameters' that are, quite often, never compared with atomic models.
Indeed, studies in the literature that even contain a
model detailed enough to connect the two scales are few and far between.

  We are also at a loss for combining models of different scales with one another.
Of the many proposed methods for coupling quantum mechanical wavefunction
calculations to continuous solvent, essentially all of them neglect explicit first-shell water structure
that could be experimentally measured with neutron scattering, diffusion measurements,
and IR and Raman spectroscopy.
Jumping directly into applications is a disease infecting much of contemporary science.
Rather than attempting to faithfully reproduce the underlying physics, many models are compared by
directly checking against experimentally measured energies -- and no clear winner has emerged (nor can it).
To be correct, models must be checked for consistency with experiments at neighboring length scales.
Similar remarks can be made for implicit solvent models coupling molecular mechanics to
continuum.
Even Marcus theory is not untouched.  There is currently debate on the proper way to
conceptualize its parameter that sets the `stiffness' of the solvent linear response.\cite{rrems15}

% Can we put the Djinn back in the bottle? [no, but we can understand its power]

  In order to make progress, we must apparently work as if we had one hand
tied behind our back.  Used correctly, simulations provide a precise tool to answer
a well-posed question within a known theory, or as a method of experimentation to discover
ideas.  However, when used absent a general theory,
simply as a tool to reproduce or predict a benchmark set of experimental data,
simulation is not capable of providing any detailed insight or understanding
of molecular science.

\section*{Acknowledgements}
I thank the anonymous reviewers for their comments and suggestions.

%\bibliography{../diel,range}

\end{document}